\documentclass[12pt]{iopart}
\usepackage[dvips]{graphicx}
\usepackage{iopams}
\begin{document}
\title{Acceleration from M theory and Fine-tuning}
\author{Yungui Gong\dag\ddag\ and Anzhong Wang\ddag}
\address{\dag\ College of Electronic
Engineering, Chongqing University of Posts and Telecommunications,
Chongqing 400065, China}
\address{\ddag\ CASPER, Physics Department, Baylor University,
Waco, TX 76798, USA} \ead{gongyg@cqupt.edu.cn}

\begin{abstract}
The compactification of M theory with time dependent hyperbolic internal space gives an
effective scalar field with exponential potential which provides a transient acceleration
in Einstein frame in four dimensions. Ordinary matter and radiation are present in addition to the
scalar field coming from compactification. We find that we have to fine-tune
the initial conditions of the scalar field so that our Universe experiences acceleration now.
During the evolution of our Universe, the volume of the internal space increases about 12 times.
The time variation of the internal space results in a large time variation of the fine structure constant
which violates the observational constraint on the variation of the fine
structure constant. The large variation of the fine structure
constant is a generic feature of transient acceleration models.
\end{abstract}
 \maketitle

\section{Introduction}
In recent years, some attempts have been made to derive a late time accelerating universe from a
fundamental theory of particle physics that incorporates gravity, like M theory or superstring theory.
This process involves a compactification of eleven or ten- dimensional Einstein gravity to four dimensions.
The no-go theorem states that we could not obtain accelerating universes in the Einstein frame in four
dimensions by compactification of pure Einstein gravity in higher dimensions when the six or seven- dimensional
internal space is a time independent nonsingular compact manifold without boundary \cite{gibbons,maldacena}.
So any attempt to derive an accelerating universe must circumvent the no-go theorem. Townsend
and Wohlfarth invoked a time dependent compactification of pure gravity in higher dimensions with hyperbolic
internal space to circumvent the no-go theorem \cite{townsend}. Their exact solution to the vacuum Einstein
gravity in higher dimensions exhibits a short period of acceleration or transient acceleration which can't explain
the early inflation but may be used to explain the late time acceleration of our universe. The solution is
the zero flux limit of spacelike branes as shown in \cite{ohta}. If non-zero flux or forms are turned on, then it
is shown that a transient acceleration exists for both compact internal hyperbolic and flat spaces
\cite{ohta,wohlfarth}. Other accelerating solutions were found by compactifying more complicated time dependent
internal spaces [6-15]. When higher order corrections were added to the higher dimensional supergravity theory,
inflationary solution was also found \cite{maeda}.

The application of time dependent compactification sheds some lights on the resolution of the late time acceleration
within a fundamental theoretical framework. However, there is no stable point in the size of the internal space
when the internal space is time dependent. This seems to be the common feature. From our four dimensional point
of view, the negatively curved internal space provides a positive exponential potential. The generic feature
of the potential is that it is too steep to sustain an eternal acceleration. In a realistic cosmological set up,
matter and radiation are present in addition to the scalar field coming from the dimensional reduction, then
the evolution of the scalar field will be different. It is worthy of studying whether the scalar field is dominant
now and it provides the late time acceleration when matter and radiation are present. Therefore, we
will study the model with matter and scalar field which comes from the time dependent compactification of M theory.
We also discuss the evolution of the fine structure constant
 due to the time dependence of the internal space. The
paper is organized as follows. In the next section, we will review the basic results of dimensional reduction.
In section 3, we will discuss the evolution of the scalar field when matter and radiation are included. In section
4, we conclude the paper with some discussions.

\section{Dimensional Reduction}
We start with the the bosonic part of the low energy effective action for M theory ($D=11$)
or superstrings ($D=10$) in $D$ dimensions:
\begin{equation}
S=\frac{1}{2\kappa^2_D}\int dx^D\sqrt{-G}\ ^{(D)}R,
\end{equation}
where we have dropped contributions from forms and dilations which are assumed to be stabilized when they exist,
$\kappa^2_D$ is the $D$ dimensional gravitational constant.
When we decompose
the $D$ dimensional space-time into $4+d$ dimensional space-time with the metric
$ds^2=\gamma_{\mu\nu}(x) dx^\mu dx^\nu+\Phi^2(x) \hat{g}_{mn}(y) dy^m dy^n$, here
the second half of the lower case Latin letters run from $1$ to $d$ and
the Greek letters run from $0$ to $3$, we get
\begin{equation}
\label{lageq1}
\fl S=\frac{1}{2\kappa^2_D}\int dx^d\sqrt{\hat{g}}\int dx^4 \sqrt{-\gamma}\Phi^d[R(\gamma)+\Phi^{-2}\hat{R}(\hat{g})
-d(d-1)\Phi^{-2}\gamma^{\mu\nu}\partial_\mu\Phi\partial_\nu\Phi-2d\Phi^{-1}\Box \Phi].
\end{equation}
After the conformal transformation
\begin{equation}
\label{conf1}
\gamma_{\mu\nu}=\Phi^{-d}g_{\mu\nu},
\end{equation}
we get the four dimensional low energy effective action in Einstein frame
\begin{equation}
\label{lageq2}
S=\frac{V_d}{2\kappa^2_D}\int dx^4 \sqrt{-g}\left[R(g)+\Phi^{-2-d}\hat{R}(\hat{g})
-\frac{d(d+2)}{2}\Phi^{-2}g^{\mu\nu}\partial_\mu\Phi\partial_\nu\Phi\right],
\end{equation}
where $V_d=\int dx^d\sqrt{\hat{g}}$ is the volume of the $d$ dimensional internal space
with the metric $\hat{g}_{mn}$ and
$\hat{R}(\hat{g})$ is assumed to be a constant. It is evident
that we can identify $\phi=\sqrt{d(d+2)/2\kappa^2}\ln(\Phi)$ as the canonical scalar field in 4 dimensions,
where $\kappa^2=8\pi G=m^{-2}_{pl}=\kappa^2_D/V_d$ is the four dimensional gravitational constant.
Since we are interested in cosmological solutions, we take
the metric of our $D$ dimensional space-time as
$ds^2=b^{-d}(t)(-dt^2+a^2(t)\tilde{g}_{ij}dx^i dx^j)+b^2(t)\hat{g}_{mn}dy^m dy^n$. The time $t$
and the scale factor $a(t)$ are the four dimensional quantities in Einstein frame. With this metric, we have
\begin{eqnarray}
^{(D)}G_{00}=3\frac{\dot{a}^2}{a^2}+\frac{3k}{a^2}-\frac{d(d+2)}{4}\frac{\dot{b}^2}{b^2}
+\frac{1}{2}b^{-2-d}\hat{R},\\
^{(D)}G_{ij}=-a^2\tilde{g}_{ij}\left[2\frac{\ddot{a}}{a}+\frac{\dot{a}^2}{a^2}+\frac{k}{a^2}
+\frac{d(d+2)}{4}\frac{\dot{b}^2}{b^2}+\frac{1}{2}b^{-2-d}\hat{R}\right],
\end{eqnarray}
\begin{equation}
\fl ^{(D)}G_{mn}=\hat{G}_{mn}(\hat{g})-3b^{d+2}\hat{g}_{mn}\left[\frac{\ddot{a}}{a}+\frac{\dot{a}^2}{a^2}+\frac{k}{a^2}
-\frac{1}{6}(d+2)\frac{\ddot{b}}{b}+\frac{(d+2)^2}{12}\frac{\dot{b}^2}{b^2}-\frac{1}{2}(d+2)\frac{\dot{a}}{a}\frac{\dot{b}}{b}\right],
\end{equation}
where $\hat{G}_{mn}(\hat{g})=\hat{R}_{mn}-\hat{g}_{mn}\hat{R}/2$
and $\hat{R}_{mn}=-(d-1)\delta^2\hat{g}_{mn}$.
The vacuum Einstein equations are
\begin{eqnarray}
\label{vaceq1}
3\frac{\dot{a}^2}{a^2}+3\frac{k}{a^2}=\frac{d(d+2)}{4}\frac{\dot{b}^2}{b^2}+\frac{1}{2}d(d-1)\delta^2 b^{-2-d},\\
\label{vaceq2}
\frac{\ddot{b}}{b}-\frac{\dot{b}^2}{b^2}+3\frac{\dot{a}}{a}\frac{\dot{b}}{b}=(d-1)\delta^2 b^{-2-d},\\
\label{vaceq3}
\frac{\ddot{a}}{a}=-\frac{d(d+2)}{6}\frac{\dot{b}^2}{b^2}+\frac{d(d-1)}{6}\delta^2 b^{-d-2}.
\end{eqnarray}
Equations (\ref{vaceq1}-\ref{vaceq3}) are not totally independent. For example, equation (\ref{vaceq3})
can be derived from equations (\ref{vaceq1}) and (\ref{vaceq2}). In the following, we will assume our
four dimensional universe is flat so that $k=0$.
To solve the above equations, we change the time coordinate $t$ to
$\eta$ so that $dt=a^3 d\eta$ and we use $'$ to denote the derivative with respect to $\eta$
in this section. By using $\eta$,
Equations (\ref{vaceq1}-\ref{vaceq3}) become
\begin{eqnarray}
\label{vaceq4}
3\frac{a^{\prime 2}}{a^2}=\frac{d(d+2)}{4}\frac{b'^2}{b^2}+\frac{1}{2}d(d-1)\delta^2 a^6 b^{-2-d},\\
\label{vaceq5}
\frac{b''}{b}-\frac{b'^2}{b^2}=(d-1)\delta^2 a^6 b^{-2-d},\\
\label{vaceq6}
\frac{a''}{a}-\frac{a^{\prime 2}}{a^2}=\frac{d(d-1)}{2}\delta^2 a^6 b^{-d-2}.
\end{eqnarray}
Let $g=a^2 b^{-d}$, then equations (\ref{vaceq5}) and (\ref{vaceq6}) give $(\ln g)''=0$, so
$a^2 b^{-d}=C_3\exp(C_4\eta)$. Take
$f=a^6 b^{-d-2}$, equations (\ref{vaceq5}) and (\ref{vaceq6}) give $(\ln f)''=2(d-1)^2\delta^2 f$, so
$a^{-6} b^{d+2}=4(d-1)^2\delta^2 C_1^2 \sinh^2[(\eta+C_2)/2C_1]$. Substitute the above solutions to equation
(\ref{vaceq4}), we get the relationship $C_1^2 C_4^2=d/3(d+2)$. Therefore, after re-scaling of $\eta$ with
the choice $C_2=0$ and $C_4=2$,
the general solution is
\begin{eqnarray}
\label{sol1}
a(\eta)=A \exp\left[
-\frac{d+2}{2(d-1)}\eta\right]\sinh^{-d/2(d+1)}\left[\sqrt{\frac{3(d+2)}{d}}|\eta|\right],\\
\label{sol2}
 b(\eta)=B \exp\left(
-\frac{3}{d-1}\eta\right)\sinh^{-1/(d-1)}\left[\sqrt{\frac{3(d+2)}{d}}|\eta|\right],
\end{eqnarray}
where $B=[d(d-1)^2\delta^2 C_3^3/3(d+2)]^{-1/2(d-1)}$ and $A=C_3^{1/2} B^{d/2}$.
So $a(\eta)\rightarrow 0$ and $b(\eta)\rightarrow \infty$ when $\eta\rightarrow -\infty$ if $d>1$.
In terms of $\eta$, the Universe starts from $-\infty$ and the infinite future time is $\eta=0$. Because
\begin{equation}
\label{expan1}
\frac{a'}{a}=-\frac{d+2}{2(d-1)}+\frac{\sqrt{3d(d+2)}}{2(d-1)}\coth\left[\sqrt{\frac{3(d+2)}{d}}|\eta|\right],
\end{equation}
so we have $\dot{a}>0$ if $d>1$, i.e., the Universe expands. To get acceleration $\ddot{a}>0$, we need to require
\begin{equation}
\label{acccond}
\frac{a'^2}{a^2}<\frac{3(d+2)}{4(d-1)}\sinh^{-2}\left[\sqrt{\frac{3(d+2)}{d}}|\eta|\right]=m(\eta).
\end{equation}
The evolution of  $m(\eta)-a'^2/a^2$ is shown in Figure 1. We see that there exists a period of acceleration.
However, the duration of acceleration is not long enough to explain the flatness and horizon problems, but
it can be used to explain the late time acceleration.
Note that
\begin{equation}
\label{expan2}
\frac{b'}{b}=-\frac{3}{d-1}+\sqrt{\frac{3(d+2)}{d(d-1)^2}}\coth\left[\sqrt{\frac{3(d+2)}{d}}|\eta|\right],
\end{equation}
it is possible that $\dot{b}>0$ when $\eta\rightarrow 0$, i.e., at late times or in the far future, the volume
of extra dimensions will increase. In the far future, $\eta\rightarrow 0$, we have
\begin{equation}
\label{ratio}
\frac{b'/b}{a'/a}\rightarrow \frac{2}{d}.
\end{equation}
Therefore, extra dimensions will manifest themselves in the future.
\begin{figure}
\centering
\includegraphics[width=12cm]{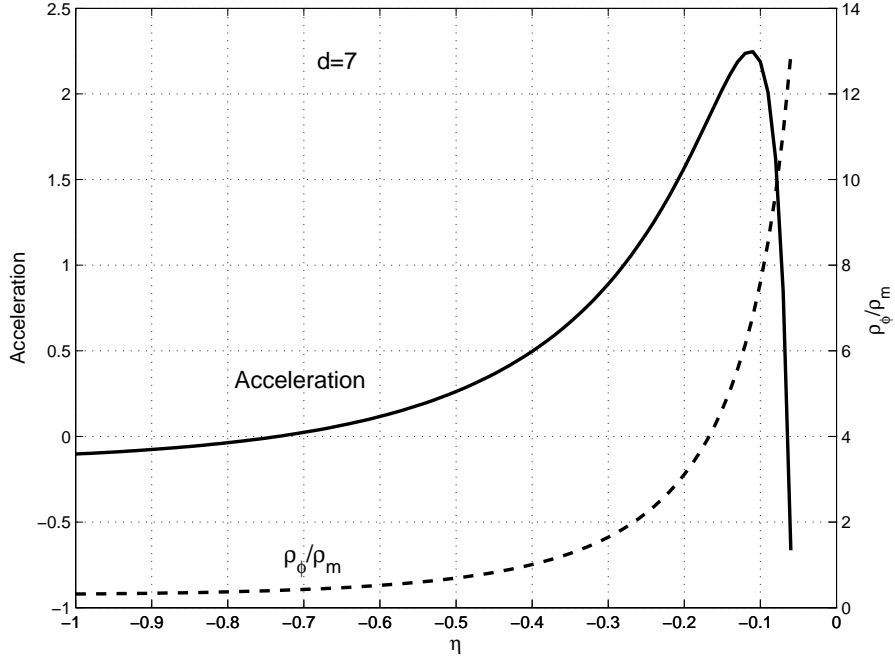}
\caption{The evolution of acceleration is plotted in the solid line by using the left $y$ axis.
The dotted line plots the evolution
of the ratio of $\rho_\phi/\rho_m$ by using the right $y$ axis.
In this plot, some constants of proportionality are not included.} \label{fig1}
\end{figure}

\section{Four Dimensional Cosmology}
From four dimensional point of view, equations (\ref{vaceq1}-\ref{vaceq3}) describe the standard cosmology
plus a scalar field $\phi=\sqrt{d(d+2)/2\kappa^2}\ln(b(t))$ with the exponential potential
$V(\phi)=[d(d-1)\delta^2/2\kappa^2]\exp(-\sqrt{2(d+2)/d}\, \phi/m_{pl})$ as the matter source.
Since we are interested in the late time acceleration instead of the early time inflation, we
should also include radiation and dust in the matter contents. With all these matter sources,
the standard cosmology is
\begin{eqnarray}
\label{eq1}
H^2=\frac{\dot{a}^2}{a^2}=\frac{1}{3m_{pl}^2}\left[\rho_m+\rho_r+\frac{1}{2}\dot{\phi}^2+V(\phi)\right],\\
\label{eq2}
\ddot{\phi}+3H\dot{\phi}+\frac{d V(\phi)}{d\phi}=0,\\
\label{eq3}
\frac{\ddot{a}}{a}=-\frac{1}{3m_{pl}^2}\left[\frac{1}{2}\rho_m+\rho_r+\dot{\phi}^2-V(\phi)\right],
\end{eqnarray}
where $\rho_m=\rho_{m0}a_0^3/a^3$ and $\rho_r=\rho_{r0}a_0^4/a^4$, the subscript $0$ means
the present value of the variable. Without dust and radiation, the Universe
with the scalar field experiences a transient acceleration as discussed in the previous section. Here we
are concerned with the question if the acceleration persists when dust and radiation are present. For
a cosmology with a barotropic fluid which has an equation of state $p_b=w_b\rho_b$ and a scalar\
field which has the exponential potential $V(\phi)=V_0\exp(-\lambda\phi/m_{pl})$, it was found that
\cite{copeland,ferreira}:

(1) $\lambda^2>3(1+w_b)$. The late time attraction solution is the tracking solution in which
the evolution of the scalar field mimics that of the barotropic fluid with $w_\phi=w_b$ and
$\Omega_\phi=3(1+w_b)/\lambda^2$, where $\Omega_\phi=\rho_\phi/3m_{pl}^2 H^2$.

(2) $\lambda^2<3(1+w_b)$. The late time attraction solution is the scalar field dominated solution
with $\Omega_\phi=1$ and $w_\phi=-1+\lambda^2/3$.

For the scalar field coming from compactification, we have $\lambda^2=2(d+2)/d<3$. So the late time
attractor solution is the scalar field dominated power-law solution and the equation of state parameter of
the scalar field is
\begin{equation}
\label{wphi}
w_\phi=-1+\frac{\lambda^2}{3}=-\frac{d-4}{3d}.
\end{equation}
If we take $d=7$, we get $w_\phi=-1/7>-1/3$. If we take $d=6$, we get $w_\phi=-1/9>-1/3$. The late
time attractor solution gives a decelerating universe. Whatever the initial conditions of the scalar
field are, the scalar field can always start from subdominant at early times and become dominant
at late times. In order for the scalar field to contribute about 70\% to the total
energy density and provide an accelerating universe at present time, it must have not reached
the fixed point yet. Therefore it is necessary to fine-tune the initial conditions of the scalar
field \cite{franca}. Before we study the fine-tuning problem, we would like to further examine the exact solutions
obtained in the previous section first. By using the exact solutions (\ref{sol1}) and (\ref{sol2}),
we get the energy density for the
scalar field $\phi$
\begin{eqnarray}
\label{endens1}
\rho_\phi&=\frac{d(d+2)}{4\kappa^2}\frac{\dot{b}^2}{b^2}+\frac{1}{2\kappa^2}d(d-1)\delta^2 b^{-d-2} \nonumber\\
&=a^{-6}\kappa^{-2}\left[\frac{d(d+2)}{4}\frac{b'^2}{b^2}+\frac{3(d+2)}{2(d-1)}\sinh^{-2}
\left(\sqrt{\frac{3(d+2)}{d}}|\eta|\right)\right].
\end{eqnarray}
Since $\rho_m\propto a^{-3}$, the evolution of $\rho_\phi/\rho_m$ is shown in Figure \ref{fig1}. During the
acceleration phase, we see that the ratio of $\rho_\phi/\rho_m$ increases with time. Therefore, we
expect the scalar field contributes more and more to the total energy density.

\begin{figure}
\centering
\includegraphics[width=12cm]{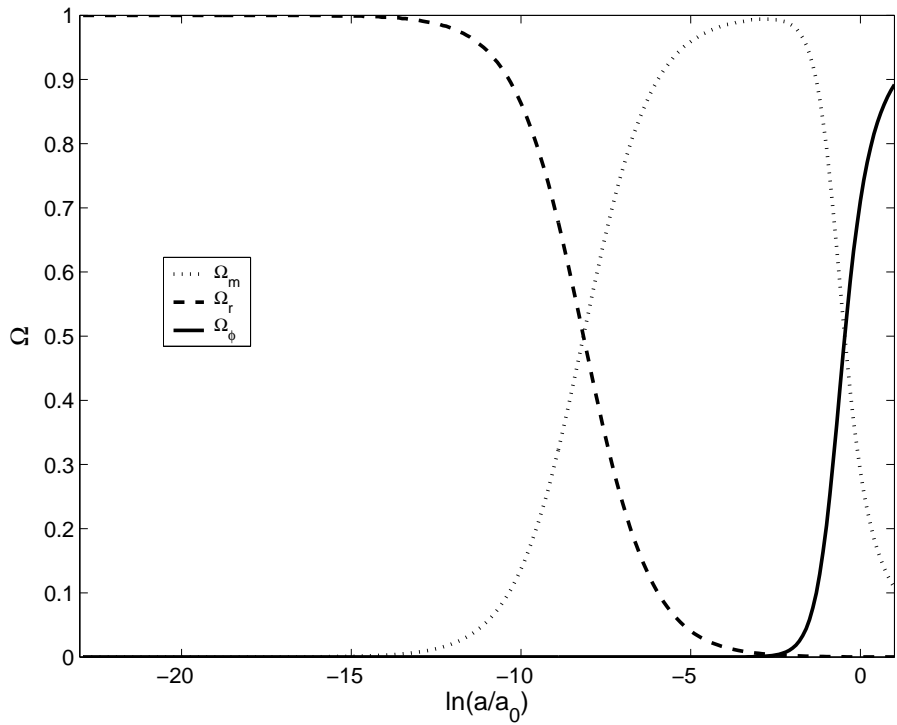}
\caption{The evolutions of the energy densities of matter, radiation and the scalar field $\phi$. The
solid line is for $\Omega_\phi$, the dashed line is for $\Omega_r$ and the dotted line
is for $\Omega_m$.} \label{fig2}
\end{figure}

\begin{figure}
\centering
\includegraphics[width=12cm]{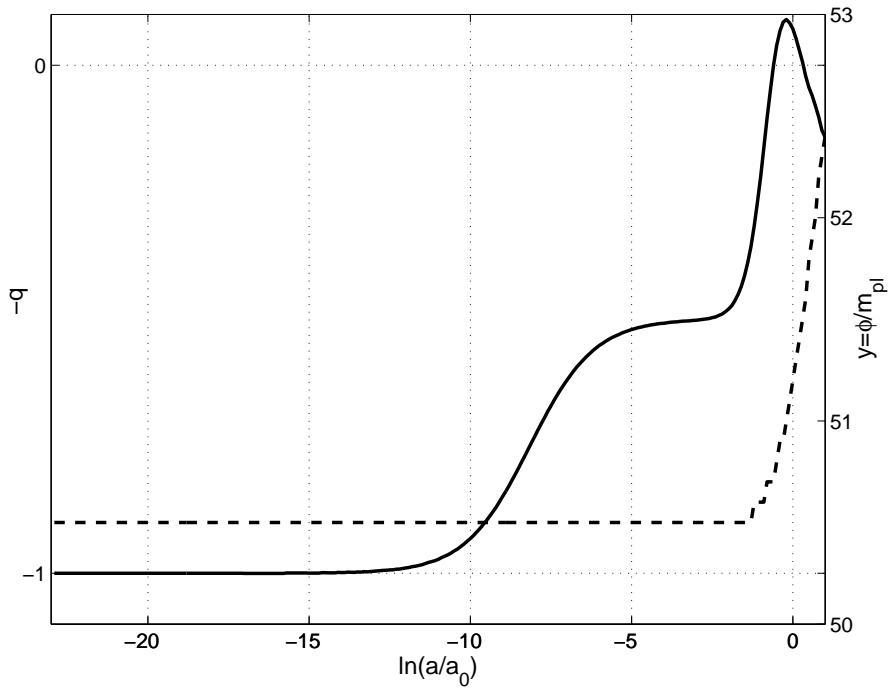}
\caption{The evolution of acceleration $\ddot{a}/aH^2$ is plotted by the solid line by using the left $y$ axis.
The dashed line plots the evolution
of the scalar field $\phi$ by using the right $y$ axis.} \label{fig3}
\end{figure}

Now we are ready to solve equations (\ref{eq1})-(\ref{eq3}) numerically. It is more convenient to use
the variable $u=\ln(a/a_0)$ instead of the cosmic time $t$. Let $y=\phi/m_{pl}$, equations (\ref{eq1})
and (\ref{eq2}) become
\begin{eqnarray}
\label{numeq1}
\left(\frac{H}{H_0}\right)^2=\frac{\Omega_{m0}\exp(-3u)+\Omega_{r0}\exp(-4u)+\Omega_{V0}\exp(-\lambda y)}{1-y'^2/6}.\\
\label{numeq2}
y''=3\lambda \Omega_V\exp(-\lambda y)-\left[\frac{3}{2}\Omega_m+\Omega_r+3\Omega_V\exp(-\lambda y)\right]y',
\end{eqnarray}
where $\Omega_m=\rho_m/3H^2m_{pl}^2=\Omega_{m0}(H_0/H)^2\exp(-3u)$,
$\Omega_V=V_1/3H^2m_{pl}^2=\Omega_{V0}(H_0/H)^2$, $V_1=d(d-1)\delta^2/2\kappa^2$,
$\Omega_\phi=y'^2/6+\Omega_V\exp(-\lambda y)$,
$\Omega_r=\rho_r/3H^2m_{pl}^2=\Omega_{r0}(H_0/H)^2\exp(-4u)$, and we use $'$ to denote
differentiation with respect to $u$ in this section. By using equation (\ref{eq3}), we get
\begin{equation}
\label{numeq3}
\frac{\ddot{a}}{a H^2}=-q=\Omega_\phi-\frac{1}{2}y'^2-\frac{1}{2}\Omega_m-\Omega_r.
\end{equation}
From equation (\ref{numeq3}), we can see that $y_0'<2\Omega_{\phi 0}$. On the other hand, $\Omega_\phi$
is small at early times, so $y'$ must also be small at early times. Therefore, $y'$ is small from the past to
the present, which means that the scalar field changes very slowly. In other words, the scalar field
behaves more like a cosmological constant so that $\Omega_{V0}\exp(-\lambda y_0)\sim 0.7$
and $\Omega_V\exp(-\lambda y)\sim 10^{-36}$ at nucleosynthesis when $a/a_0\sim 10^{-10}$.
From equation (\ref{numeq2}), we see that at very early time when radiation was dominated,
$y''\sim -y'$. So for small positive $y'$, $y'$ will decrease first until it reaches below $10^{-36}$,
then it will start to increase again. That is another way to see why the scalar field changes very slowly.
In the recent past, we have $y''\sim 3 \lambda \Omega_\phi$, so $y'$ became relatively large
and the scalar field would change relatively fast in the
recent past. For the numerical calculation, we start
from the epoch of nucleosyntheis around $a/a_0\sim 10^{-10}$ or $u=-23$. The initial conditions are chosen
as $\Omega_{Vi}\exp(-\lambda y_i)\sim 10^{-36}$ and $y_i'=0.01$. The evolution of $\Omega_\phi$, $\Omega_m$
and $\Omega_r$ are shown in Figure \ref{fig2}.
With the choice of the initial conditions of the scalar field, the behaviors of the energy densities
are consistent with the current observations. We also plot the evolutions of the dimensionless acceleration
$-q=\ddot{a}/aH^2$ and the scalar field in Figure \ref{fig3}.
From Figure \ref{fig3}, we see that the scalar field increases very slowly for most of the time and it starts
to increase relatively fast in the recent past when it became to dominate the energy component. This behavior
confirmed the above discussions. $\phi/m_{pl}$ increases about $2$ during the whole period of evolution, so
the volume increases about $\exp(2\sqrt{2d/(d+2)})=12.1$ times.

\section{Discussions}
Upon compactification of pure gravity in higher dimensions with time dependent hyperbolic
internal space, we get a scalar field with exponential potential in four dimensions. The potential
of the scalar field is too steep to give an accelerated attractor solution. With suitable
fine-tuning of the initial conditions of the scalar field, our Universe experiences a transient
acceleration now. The scalar field changed very slowly until recently and it behaves like
a cosmological constant. Due to the scalar field dynamics, the volume of the internal space increases
about $12.1$ times. In the far future, the internal space will become infinite and $(\dot{b}/b)/(\dot{a}/a)$
goes to $2/d$ asymptotically. The change of the volume of the internal space results in the variation
of the fundamental constants. Cline and Vinet found that the time variation of the gravitational
constant violated the observational constraint in the string frame for time-varying extra dimensions \cite{cline}.
It is known that the effective four dimensional fine structure constant is inversely proportional to the volume
of the internal space, so
\begin{equation}
\label{fine}
\left|\frac{\dot{\alpha}}{\alpha}\right|=d \left|\frac{\dot{b}}{b}\right|=\sqrt{\frac{2d}{d+2}} H |y'|.
\end{equation}
The experimental data gives $|\dot{\alpha}/\alpha|\lesssim 10^{-14}$ yr$^{-1}$ \cite{webb}.
Substitute this constraint and $H_0\sim 10^{-10}$ to equation (\ref{fine}), we get
\begin{equation}
\label{bound}
|y'|\lesssim 10^{-5}.
\end{equation}
As we discussed in the last section, the scalar field changes relatively fast during the acceleration
phase and the bound (\ref{bound}) is violated at present. Although the time dependent compactification of
M theory or superstring theory can give a transient acceleration which is consistent with the current
astronomical observations, the time variation of the internal space will result in large variation
of the fine structure constant. This result is quite general and can apply to other models
with transient acceleration. The change of fine structure constant due to time dependent internal space
was also discussed in \cite{baukh}. They discussed the scalar field dominated region only. In our
discussion, we include radiation and cold dark matter, so our result is more general.

\ack
Y. Gong is supported by Baylor University, NNSFC under grant No. 10447008 and 10575140,
SRF for ROCS, State Education Ministry
and CQUPT under grant No. A2004-05.


\section*{References}
\begin{thebibliography}{30}
\bibitem{gibbons} Gibbons G W 1985 {\it Supersymmetry, Supergravity and Related Topics} ed de Aguila F,
de Azc\'{a}rraga and Iba\~{n}ez (Sigapore: World Scientific) p 124
\bibitem{maldacena} Maldacena J M and Nu\~{n}ez C 2001 {\it Int. J. Mod. Phys.} A {\bf 16} 822
\bibitem{townsend} Townsend P K and Wohlfarth N R 2003 \PRL {\bf 91} 061302
\bibitem{ohta} Ohta N 2003 \PRL {\bf 91} 061303
\bibitem{wohlfarth} Wohlfarth N R 2003 \PL B {\bf 563} 1
\nonum
Roy S 2003 \PL B {\bf 567} 322
\bibitem{neupane1} Neupane I P and Wiltshire D L 2005 \PL B {\bf 619} 201
\bibitem{neupane2} Neupane I P and Wiltshire D L 2005 \PR D {\bf 72} 083509
\bibitem{chen} Chen C M {\it et al} 2003 {\it J. High Energy Phys.} JHEP10(2003)058
\bibitem{bergshoeff} Bergshoeff E 2004 \CQG {\bf 21} 1947
\bibitem{collinucci} Collinucci A, Nielsen M and Van Riet T 2005 \CQG {\bf 22} 1269
\bibitem{jarv} Jarv L, Mohaupt and Saueressig 2004 {\it J. Cosmol. Astropart. Phys.} JCAP02(2004)012
\bibitem{ohta1} Ohta N 2005 {\it Int. J. Mod. Phys.} A {\bf 20} 1
\bibitem{ohta2} Ohta N 2003 {\it Prog. Theor. Phys.} {\bf 110} 269
\bibitem{baukh} Baukh V and Zhuk A 2006 {\it Preprint} hep-th/0601205
\bibitem{saffin} Karthauser J L P and Saffin P M 2006 {\it Preprint} hep-th/0601230
\bibitem{maeda} Maeda K and Ohta N 2005 \PR D {\bf 71} 063520
\bibitem{copeland} Copeland E J, Liddle A R and Wands D 1998 \PR D {\bf 57}, 4686
\bibitem{ferreira} Ferreira P G and Joyce M 1997 \PRL {\bf 79}, 4740
\bibitem{franca} Fran\c{c}a U and Rosenfeld R 2002 {\it J. High Energy Phys.} JHEP10(2002)015
\bibitem{cline} Cline J M and Vinet J 2003 \PR D {\bf 68} 025015
\bibitem{webb} Webb J K {\it et al} 2001 \PRL {\bf 87} 091301
\end{thebibliography}
\end{document}